\documentclass[epj]{svjour}

\usepackage{graphicx}
\usepackage{fancyhdr}
\usepackage{amsmath,amssymb,psfrag}
\def\makeheadbox{\flushright{{HIP-2007-52/TH}\\{DO-TH 07/09}}}
\def\LHC{{\small LHC}}
\def\SUSY{{\small SUSY}}
\def\WMAP{{\small WMAP}}
\def\hc{\textrm{\scriptsize H.C.}}
\def\stilde{\widetilde}
\def\tanb{\ensuremath{\tan\beta}}
\def\vev{\emph{vev}}

\def\mytitle{My title} 
\def\myauthors{My name}  
\def\mytype{My type of session}
\def\mysession{My session}

\def\mytitle{Non-universal gaugino masses and implications on the dark
  matter and Higgs searches} 
\def\myauthors{Jari Laamanen}  
\def\mytype{Contributed Talk}    
\def\mysession{Cosmology and Astrophysics}

\begin{document}
\title{Non-universal gaugino masses and implications on the dark
  matter and Higgs searches} 
\author{Katri Huitu\inst{1}
  \and 
  \underline{Jari Laamanen}
  \inst{2}
  \fnmsep
  \inst{1}
  \fnmsep
  \thanks{\emph{Email:} jari.laamanen@uni-dortmund.de}%
  \and Sourov Roy\inst{3}
}
\institute{Helsinki Institute of Physics 
  and High Energy Physics Division, 
  Department of Physical
  Sciences, PL 64, {FIN-00014},
  University of Helsinki, Finland
  \and
  Institut f{\"u}r Physik, 
  Universit{\"a}t Dortmund, 
  D-44221 Dortmund, Germany
  \and 
  Department of Theoretical Physics 
  and Centre for Theoretical Sciences,
  Indian Association for the Cultivation of Science, 
  2A \& 2B Raja S.C. Mullick Road, 
  Kolkata 700 032, India
}
%
\date{}
\abstract{Non-universal boundary conditions in grand unified theories
  can lead to non-universal gaugino masses at the unification scale.
  In $R$-parity preserving theories the lightest supersymmetric
  particle is a natural candidate for the dark matter. We have studied
  the composition of the lightest neutralino in non-universal gaugino
  mass cases from the SU(5), and implications on the dark matter. In
  the representations of SU(5) thermal relic density agreeing with
  \WMAP\ is found. The possibility to observe the neutral MSSM Higgs
  bosons ($h/H/A$) at the \LHC\ via neutralino cascades when the
  lightest neutralino is dark matter, is discussed for the
  representation {24}, and the connection to dark matter is
  established.
} 
\maketitle

\section{Introduction}
\label{intro}
Most of the phenomenological studies involving neutralinos have been
performed with universal gaugino masses at the grand unification
scale. However, there is no compelling theoretical reason for such a
choice. This talk is based on the reference \cite{parent}, where the
non-universal gauginos arising from SU(5) are considered.

In grand unified supersymmetric models, which include an ${\rm SU(5)}$
grand unified model, non-universal gaugino masses are generated by a
nonsinglet chiral superfield $\Phi^n$ that appears linearly in the
gauge kinetic function $f(\Phi)$
\cite{Ellis:1985jn,Drees:1985bx}. The function
$f(\Phi)$ is an analytic function of the chiral superfields $\Phi$ in
the theory. It should be noted that the chiral superfields $\Phi$
consist of a set of gauge singlet superfields $\Phi^s$ and gauge
nonsinglet superfields $\Phi^n$, respectively, under the grand unified
group. If the auxiliary part $F_\Phi$ of a chiral superfield $\Phi$ in
the $f(\Phi)$ gets a \vev, then gaugino masses arise from the coupling
of $f(\Phi)$ with the field strength superfield $W^a$. The Lagrangian
for the coupling of gauge kinetic function with the gauge field
strength is written as
\begin{eqnarray}
  {\cal L}_{gk} = \int d^2 \theta f_{ab}(\Phi) W^a W^b + \hc,
  \label{gauge-kinetic}
\end{eqnarray} 
where a and b are gauge group indices [for example, a,b = 1,2,...,24
for ${\rm SU(5)}$], and repeated indices are summed over. The gauge
kinetic function $f_{ab}(\Phi)$ is
\begin{eqnarray}
  f_{ab}(\Phi) = f_0(\Phi^s)\delta_{ab} + \sum_n f_n(\Phi^s)
   \frac{\Phi^n_{ab}}{M_P} + \cdots,
\end{eqnarray}
where $\Phi^s$ and $\Phi^n$ are the singlet and nonsinglet chiral
superfields, respectively. Here $f_0(\Phi^s)$ and $f_n(\Phi^s)$ are
functions of gauge singlet superfields $\Phi^s$, and $M_P$ is some
large scale. When $F_\Phi$ gets a \vev\ $\langle F_\Phi \rangle$, the
interaction (\ref{gauge-kinetic}) gives rise to gaugino masses$\colon$
\begin{eqnarray}
  {\cal L}_{gk} \supset \frac{{\langle F_\Phi \rangle}_{ab}}
   {M_P}\lambda^a \lambda^b + \hc,
\end{eqnarray}
where $\lambda^{a,b}$ are gaugino fields. The ${\rm U(1)}$, ${\rm
  SU(2)}$, and ${\rm SU(3)}$ gauginos are denoted by $\lambda_1$,
$\lambda_2$, and $\lambda_3$, respectively.

Since the gauginos belong to the adjoint representation of the gauge
group, in the case of ${\rm SU(5)}$ for example, $\Phi$ and $F_\Phi$
can belong to any of the following representations appearing in the
symmetric product of the two {\bf 24} dimensional representations of
${\rm SU(5)}\colon$
\begin{eqnarray}
  {({\mathbf {24}} \otimes {\mathbf {24}})}_{Symm} = {\mathbf 1} \oplus
  {\mathbf {24}} \oplus {\mathbf {75}} \oplus {\mathbf {200}}.
\end{eqnarray}
In the minimal case (which is the simplest one too), $\Phi$ and
$F_\Phi$ are assumed to be in the singlet representation of ${\rm
  SU(5)}$. This corresponds to equal gaugino masses at the grand
unified theory (GUT) scale.  However, $\Phi$ can belong to any of the
nonsinglet representations ${\mathbf {24}}$, ${\mathbf {75}}$, and
${\mathbf {200}}$ of ${\rm SU(5)}$. In that case, the gaugino masses
are unequal but related to one another via the representation
invariants. It should be kept in mind that an arbitrary combination of
these different representations is also allowed but we shall study the
case of each representation separately. As we shall discuss later, the
${\mathbf {24}}$ dimensional representation is the most interesting
one in the context of our present investigation.  In Table
\ref{tab:gaug} we display the ratios of resulting gaugino masses at
tree level as they arise when $F_\Phi$ belongs to various
representations of ${\rm SU(5)}$. Clearly, the nonsinglet
representations have characteristic mass relationships for the
gauginos at the GUT scale. The resulting relations at the electroweak
scale, using the renormalization group evolution at the one-loop level
are also displayed.
\begin{table}
  \centering
  \vskip 0.1 in
  \caption{Ratios of gaugino masses at the GUT scale in the normalization 
    $M_3 ({\rm GUT})$ = 1 and at the electroweak scale in the normalization 
    $M_3 ({\rm EW})$ = 1 at the one-loop level.}
  \begin{tabular}{|l|ccc|rrc|}
    \hline \hline
    $F_\Phi$  &  $M^G_1$ & $M^G_2$ & $M^G_3$ & 
    $M^{\rm EW}_1$ & $M^{\rm EW}_2$ & $M^{\rm EW}_3$ \\
    \hline
    $\mathbf{1}$   & 1 & 1 & 1 &0.14 & 0.29 & 1\\
    $\mathbf{24}$  &--0.5 &--1.5 &1 &--0.07 & --0.43 & 1\\
    $\mathbf{75}$  &--5  &3  &1  &--0.72 & 0.87 & 1\\
    $\mathbf{200}$ &10  &2  &1  &1.44 & 0.58 & 1\\
    \hline
    \hline
  \end{tabular}
  \label{tab:gaug}
\end{table}

The phenomenology of supersymmetric models depends crucially on the
compositions of neutralinos and charginos.  In addition to the
laboratory studies, relevant input is obtained from the dark matter
searches.  The \WMAP\ satellite has put precise limits on the relic
density. Supersymmetric theories which preserve $R$-parity contain a
natural candidate for the cold dark matter particle.  If the lightest
neutralino is the lightest supersymmetric particle (LSP), it can
provide the appropriate relic density.

\section{Dark matter in SU(5) representations}
\label{sec:rd}

In many supergravity type models the lightest neutralino is bino-like,
which often leads to too high thermal relic density, as compared to
the limits provided by the \WMAP\ experiment \cite{Spergel:2006hy}.
The non-universal gaugino masses change this considerably. When the
gaugino masses are not universal at the GUT scale, the resulting
neutralino composition changes from the usual universal gaugino mass
case \cite{Huitu:2005wh}.

\subsection{Representation 1}
\label{subsec:rep1}

\begin{figure}
  \centering
  \psfrag{m2}{\hspace{-2mm}$M_{2} (= m_{1/2})$}
  \psfrag{m0}{\large $m_{0}$}
  \psfrag{lsp}{\hspace{-8.5mm} \footnotesize $\chi^0\! \neq$ \textsf{lsp}}
  \includegraphics[width=0.45\textwidth]{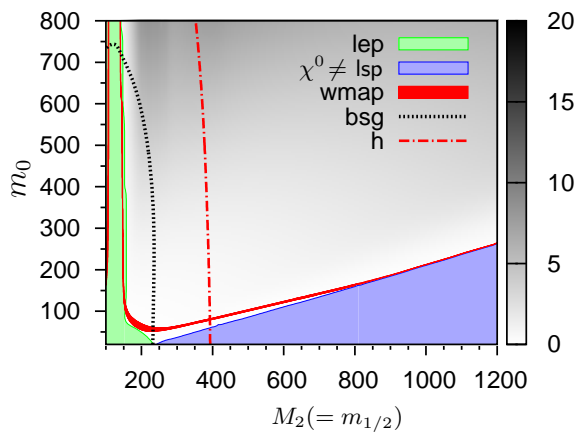}
  \caption{Singlet: $\tan\beta = 10,\ \mathrm{sgn} ( \mu ) = +1,\ A_0=0$}
  \label{fig:relic1}
\end{figure}

In Fig.~\ref{fig:relic1} the area of preferred thermal relic density
in the representation \textbf{1} is plotted for a set of (GUT scale)
parameters for the reference.
For the chosen parameters the \WMAP\ preferred regions are found near
the $M_2$ (\emph{i.e.} $m_{1/2}$ in universal gaugino mass language)
and $m_0$ axes.
The dark shaded areas represent larger relic density than the lighter
areas.  \textsf{wmap} denoted filling is the \WMAP\ preferred region,
\textsf{lep} shows an area, where the experimental mass limits are not
met, \textsf{rge} shows an area where there is no radiative EWSB, and
\textsf{lsp} the area where neutralino is not the LSP. For the relic
density, we use here the \WMAP\ combined three year limits
\cite{Spergel:2006hy}
\begin{eqnarray}
  \Omega_{CDM} h^2 = 0.11054^{+0.00976}_{-0.00956} \quad (2\sigma).
\end{eqnarray}
The area enclosed by the \textsf{bsg} contour is disallowed by $b \to
s \gamma$ constraint.
For the particle masses, the following limits are applied
\cite{Belanger:2006is}: $m_{\tilde e_R} > 99.4$ or 100.5 GeV depending
if the lightest neutralino mass is below or above 40 GeV, $m_{\tilde
  \mu_R} > 95$ GeV, $m_{\tilde\tau_1} > 80.5$ to 88 GeV depending on
the lightest neutralino mass (from 10 to 75 GeV), $m_{\tilde \nu_i} >
43$ GeV, and $m_{\tilde\chi^\pm} > 73.1$ to 103 GeV depending on the
sneutrino masses (from 45 to 425 GeV).  The curve $m_h=114$ GeV is
depicted in the figure (line denoted by \textsf{h}).  For the shown
parameter region, when otherwise experimentally allowed, Higgs is
always heavier than $91$ GeV, which is the Higgs mass limit in MSSM
for $\tan\beta \ge 10$ assuming maximal top mixing
\cite{unknown:2001xy}.
We have used the two sigma world average of
$BR(b\to s \gamma) = (355 \pm 24^{+9}_{-10} \pm 3) \times 10^{-6}$
for the branching fraction \cite{Barberio:2007cr}.
The preferred relic density area is quite constrained, and often the
neutralino relic density is overclosing the Universe.

\subsection{Representation 24 -- large relic density}
\label{subsec:rep24}
\begin{figure}[h]
  \centering
  \psfrag{m2}{$M_{2}$}
  \psfrag{m0}{\large $m_{0}$}
  \psfrag{lsp}{\hspace{-6.6mm} \footnotesize $\chi^0\!\! \neq \! \textsf{lsp}$}
  \includegraphics[width=0.45\textwidth]{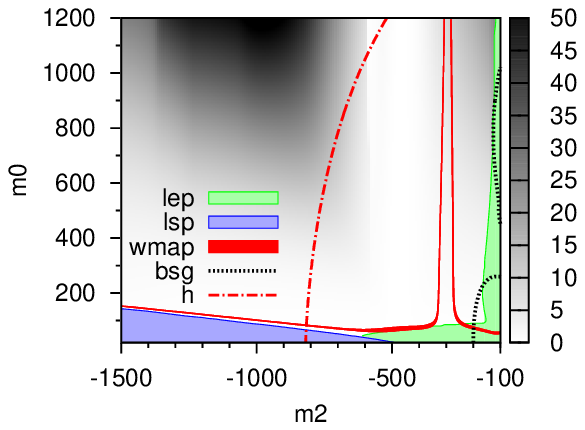}
  \caption{\textbf{Rep 24}: $\tan\beta = 10,\ \mathrm{sgn} ( \mu ) = +1,\ A_0=0$}
  \label{fig:relic24}
\end{figure}

The amount of thermal relic density in the representation \textbf{24}
is presented in Fig.~\ref{fig:relic24} for a set of GUT scale
parameters.
The most striking feature in Fig.~\ref{fig:relic24} is the valley of
the low relic density area around $M_2 \sim -300$ GeV. The minimum
occurs at the $Z$ peak providing an efficient annihilation of the
neutralinos to quarks.  Outside of the valley the relic density rises,
overclosing the universe.
In the representation \textbf{24} the lightest neutralino is very
bino-like, and the \WMAP\ preferred region tends to be quite narrow.
Many values of $m_0$ are allowed, but only for specified $m_2$. The
Higgs mass is always greater than 91 GeV for these parameters.
Higgsino component can be increased by increasing \tanb. This results
in a larger higgsino component in the lightest neutralino, which then
annihilates more efficiently.

\subsection{Representation 75 -- large higgsino component}
\label{subsec:rep75}
\begin{figure}
  \centering
  \psfrag{m2}{$M_{2}$}
  \psfrag{m0}{\large $m_{0}$}
  \psfrag{lsp}{\hspace{-7.5mm} \footnotesize $\chi^0\!\! \neq \! \textsf{lsp}$}
  \includegraphics[width=0.45\textwidth]{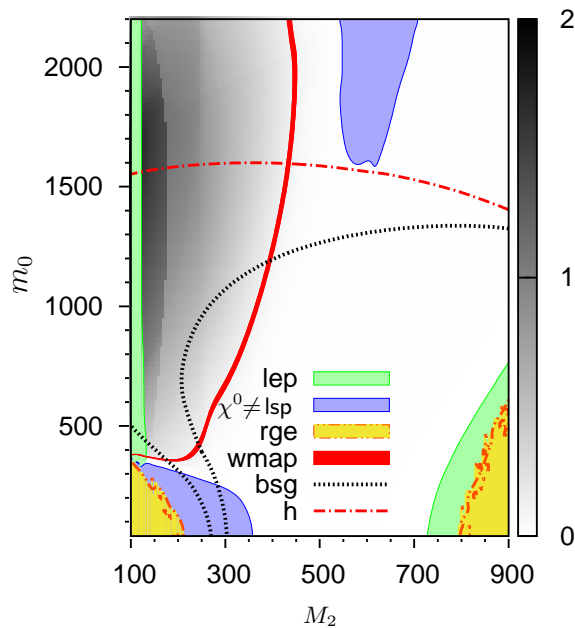}
  \caption{\textbf{Rep 75} $\tan\beta = 10,\ \mathrm{sgn}(\mu) = -1,\ A_0 = 1$ TeV}
  \label{fig:relic75}
\end{figure}

In Fig.~\ref{fig:relic75} the area of preferred thermal relic density
in the representation \textbf{75} is plotted for one set of (GUT
scale) parameters.
Since the higgsino component in the representation \textbf{75} is
large \cite{Huitu:2005wh}, the resulting relic density is low, and
most of the parameter space is not overclosed by the \WMAP\ limits.
Also the co-annihilations with the lightest chargino reduces the relic
density, since the lightest neutralino and chargino are nearly mass
degenerate in the higher $M_2$ part of the parameter space.
This is also seen in the Fig.~\ref{fig:relic75} at high $m_0$, where
the lightest chargino becomes the LSP for specific $M_2$ values.
At the low $m_0$ region the LSP can be the lighter stop.
In the low $M_2$ region the EWSB condition pushes the value of $\mu$
high, which in turn decreases the higgsino component in the lightest
neutralino, making it mostly a bino. The lightest neutralino and
chargino are not degenerate anymore, and the relic density increases
in low $M_2$ area. This enables the emergence of the \WMAP\ preferred
region in the parameter space. The second lightest neutralino can
annihilate also directly into gauge bosons in this parameter region.
Again, increasing $\tan \beta$ enhances the higgsino component leading
to lower relic densities in general.

\subsection{Note for representation 200}
\label{subsec:rep200}
In the representation \textbf{200} the lightest neutralino and
chargino are almost degenerate, so the co-annihilations reduce the
relic density substantially.  Also the Higgsino mixing is large, and,
more importantly, bino component is very small. Therefore, the
resulting relic density is tiny.  In contrast to the \textbf{75}
dimensional case, the value of the $\mu$ parameter decreases with
decreasing $M_2$, so the bino component does not get very large.
The parameter space suitable for finding partial neutralino dark
matter can be extended both in $m_0$ and in $M_2$ beyond 1 TeV for
both signs of $\mu$, but neutralino in this representation can never
be the only source of dark matter.

It is also possible to have contributions from many representations
simultaneously. In any point of the parameter space, \WMAP\ limits can
be reached by suitably combining representations. For an example, see
\cite{parent}.

\section{Higgses from cascades}
\label{sec:higgs-cascades}
If the squarks and gluinos are light enough, their production cross
sections are large at the \LHC. The light neutralinos
$\tilde{\chi}_{1,2}$ are typical decay products of $\tilde{g}$ and
$\tilde{q}$. The neutral Higgs bosons can be produced in the decay of
$\tilde{\chi}_{2}$, if the mass difference between $\tilde{\chi}_{2}$
and $\tilde{\chi}_{1}$ is large enough.  As the production rate is
largely independent on the value of tan$\beta$, these production
channels have been found particularly interesting at the \LHC\ to cover
the difficult region of low and medium tan$\beta$ values \cite{Datta:2003iz}.

\begin{figure}
  \psfrag{mg}{\large $m_{\tilde g}$}
  \psfrag{ma}{$m_{A}$}
  \psfrag{mA}{$m_{A}$}
  \psfrag{tanb}{\large $\tan\beta$}
  \psfrag{H}{\hspace{-10mm}$\sigma( \to H\tilde{\chi}_1^0 )$ \small [pb]}
  \includegraphics[width=0.45\textwidth]{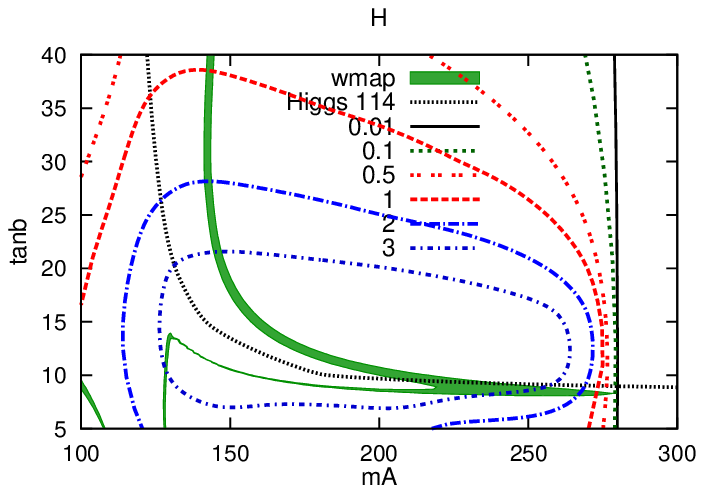}\\
  \includegraphics[width=0.45\textwidth]{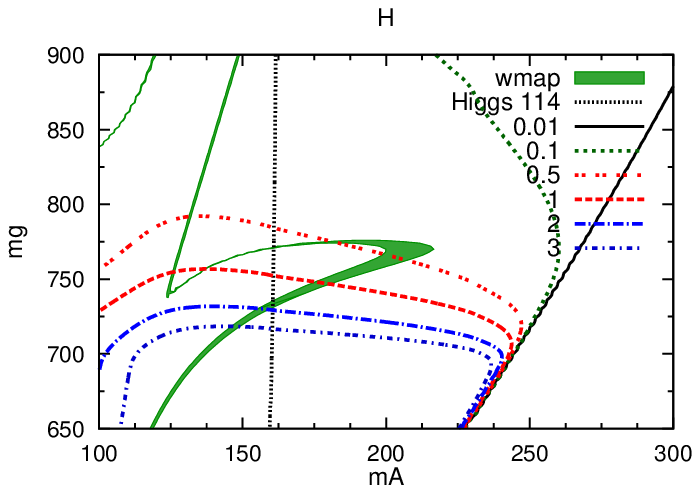}
  \caption{Cross-section of the chain $pp \to \tilde{\chi}_2^0 + X \to
    \tilde{\chi}_1^0 H + X \to \tilde{\chi}_1^0 b \bar b + X$ in the
    representation 24}
  \label{fig:cs24}
\end{figure}

A possible way to look for the Higgs bosons is through the cascade
\begin{multline}
  \label{eq:cascade}
  pp
  \to \tilde q \tilde q/ \tilde q \bar {\tilde q}/ 
  \tilde q \tilde g/ \tilde g \tilde g
  \to { \tilde{\chi}_2^0} + X \\
  \to { \tilde{\chi}_1^0}\, { h}/{ H}/{ A} +X
  \to \tilde{\chi}_1^0
  b\bar b + X.
\end{multline}
%
In Fig.~\ref{fig:cs24} the cross section for the process $pp \to H +
\tilde{\chi}_1^0 + X \to b \bar b + \tilde{\chi}_1^0 + X$ in the
$(m_A, \tan\beta)$ (upper) and $(m_A, m_{\tilde g})$ (lower) planes
for the heavier neutral Higgs scalar are plotted.  The solid (green)
fill denotes the \WMAP\ preferred relic density region.  Also the
Higgs 114 GeV mass contour is plotted with the contours of constant
cross section (larger $m_A$ values correspond to larger $m_h$).
The values of the parameters are $\mu = +700$ GeV, $m_{\tilde{q}} =
600$ GeV, $m_{\tilde{\ell}} = 350$ GeV, and the trilinear coupling for
the top sector is chosen to be $A_t = 800$ GeV. In the $(m_A,
\tan\beta)$ figure the gluino mass is chosen as $m_{\tilde{g}} = 770$
GeV, and in the $(m_A, m_{\tilde g})$ plane $\tan\beta = 10$.
The choice of $m_{\tilde g}>m_{\tilde q}$ means that every gluino
decays to a quark and the corresponding squark pair ($q \stilde q$).

For the three neutral scalar Higgses, the cross section is largest for
the $H$ Higgs production.  The stripe of \WMAP\ preferred relic
density passes through the large cross section area. For the $A$
production, the cross section is somewhat smaller, while for the $h$
production it is substantially smaller for the \WMAP\ preferred area.
In the $(m_A, \tan\beta)$ plane the near-horizontal \WMAP\ stripe
around $\tan\beta = 8$ (beginning at $m_A = 130$ GeV, $\tan\beta =
13$) corresponds to the closing of the Higgs resonance in the LSP
annihilation: in the region below that line the lightest neutralino
mass is more than half of the light Higgs boson mass. Therefore the
annihilation never occurs at the resonance.
Heavier $A$ corresponds to the larger relic density. The relic density
is lowest in between the two relic density stripes, the minimum
occurring at the $A$-peak.

In the $(m_A, m_{\tilde g})$ plane the large $H$ production cross
section is again passed by the preferred relic density stripe.  The
Higgs 114 GeV limit divides the parameter space in two (vertical line
in the lower plot of Fig.~\ref{fig:cs24}; it should be kept in mind,
though, that for the $\tan\beta = 10$ used here the actual Higgs boson
mass limit can be as low as around 90 GeV).  Also here the relic
density is the lowest in between the two relic density stripes.  The
horizontal kink in the relic density stripe around $m_{\tilde g} \sim
770$ GeV corresponds to the closing of the Higgs resonance in the LSP
annihilation: in contrast to the $(m_A, \tan\beta)$-figure, the Higgs
resonance is open below that line due to the decreasing of the
lightest neutralino mass.
%
The $A$ production cross section is slightly
smaller than for $H$, while $h$ production is the one with the lowest
cross section.
For comparison, representation {\bf 1} gives only the lightest Higgs
channel.

\section{Summary}
\label{sec:summary}
We studied the dark matter allowed regions in the SU(5) GUT
representations, of which all but the singlet lead to non-universal
gaugino masses.  The \WMAP\ preferred relic density regions were quite
distinct for different representations, which leads to a suggestion
that combinations of different representations can give observed dark
matter for otherwise experimentally allowed parameter values.

Production of the neutral Higgs bosons in the \SUSY\ cascades with
$b\overline{b}$ decay modes was studied.  In the {\bf 24} dimensional
representation the main decay modes are to heavy Higgs bosons, whereas
in the {\bf 1} dimensional representation only the lighter scalar
channel is kinematically possible (with the parameter choice we have
used).
We especially concentrated on the representation {\bf 24}, since there
the Higgs signal from the neutralino decay is interesting and
different from the usual universal singlet model.  It is important to
realize that there is no automatically theoretical preference for the
gaugino masses to be unified.
\begin{acknowledgement}
  This work was supported by the Academy of Finland (project 115032).
  The work of J.L.~is partially supported by the Bundesministerium
  f\"ur Bildung und Forschung, Berlin-Bonn.
\end{acknowledgement}

\bibliographystyle{unsrt}
\bibliography{laamanen}

\end{document}